\documentclass[lettersize,journal]{IEEEtran}
\usepackage{amsmath,amsfonts}
\usepackage[ruled]{algorithm2e}
\usepackage{array}
\usepackage[caption=false,font=normalsize,labelfont=sf,textfont=sf]{subfig}
\usepackage{textcomp}
\usepackage{stfloats}
\usepackage{url,lipsum}
\usepackage{verbatim}
\usepackage{graphicx}
\usepackage{cite}
\usepackage{mathrsfs}
\usepackage{amsmath}
\usepackage{booktabs}
\usepackage{multirow,url}
\usepackage{bm}
\usepackage[table]{xcolor}
\usepackage{bbding}
\usepackage{diagbox}
\usepackage{arydshln}
\usepackage{color}  
\def\BibTeX{{\rm B\kern-.05em{\sc i\kern-.025em b}\kern-.08em
    T\kern-.1667em\lower.7ex\hbox{E}\kern-.125emX}}

\def\bw{{\mathbf w}}
\def\bs{{\mathbf s}}
\def\bg{{\mathbf g}}

\def\cred{\textcolor{red}}

\usepackage{balance}
\begin{document}
\title{Distributed Speech Dereverberation Using Weighted Prediction Error}
\author{Ziye Yang,~\IEEEmembership{Student Member,~IEEE}, Mengfei Zhang,~\IEEEmembership{Student Member,~IEEE},   \,Jie Chen,~\IEEEmembership{Senior Member,~IEEE}
\thanks{The authors are with Northwestern Polytechnical University, China. Corresponding email: dr.jie.chen@ieee.org}
\vspace{-0.5cm}
}


\markboth{Journal of \LaTeX\ Class Files,~Vol.~18, No.~9, September~2020}%
{How to Use the IEEEtran \LaTeX \ Templates}

\maketitle

\begin{abstract}
Speech dereverberation aims to alleviate the negative impact of late reverberant reflections. The weighted prediction error (WPE) method is a well-established technique known for its superior performance in dereverberation. However, in scenarios where microphone nodes are dispersed, the centralized approach of the WPE method requires aggregating all observations for inverse filtering, resulting in {a} significant computational burden.  This paper introduces a distributed speech dereverberation method that emphasizes low computational complexity at each node. Specifically, we leverage the distributed adaptive node-specific signal estimation (DANSE) algorithm within the multichannel linear prediction (MCLP) process. This approach empowers each node to perform local operations with reduced complexity while achieving {the} global performance through inter-node cooperation. Experimental results validate the effectiveness of our proposed method, showcasing its ability to achieve efficient speech dereverberation in dispersed microphone node scenarios.
\end{abstract}


\section{Introduction}\label{sec:intro}
In enclosed spaces, speech signals captured by far-field microphone arrays often contain reverberant components caused by reflections off objects within the room. This  phenomenon can significantly degrade the quality of the target speech signal, presenting challenges for various back-end speech applications, such as automatic speech recognition systems~\cite{chetupalli2019late}. Speech dereverberation, a research area that has received substantial attention over the past few decades~\cite{kodrasi2016joint, braun2018evaluation, huang2020simple} aims to eliminate late reflections from the received reverberant speech, thereby enhancing overall speech intelligibility and quality.
Among the numerous classical speech dereverberation techniques, {multichannel linear prediction}~(MCLP) methods notably stand out~\cite{nakatani2008speech, nakatani2010speech}, particularly under conditions characterized by long acoustic impulse responses and extended reverberation time. Building on this, the {weighted prediction error}~(WPE) method~\cite{nakatani2010speech} demonstrates promising performance. This method assembles observations from all microphones on a reference channel to estimate and subtract late-reverberant components from the reverberant speech, epitomizing a typical approach to speech dereverberation.

In recent smart home or immersive online conference applications,  microphones may be dispersively deployed in a room with each microphone associated with a computational unit of limited capacity. While the centralized strategy
has shown significant potential, the process of gathering and processing all available signals on a reference channel poses challenges in terms of communication bandwidth and computational complexity. To address these limitations, researchers have focused on developing distributed estimation techniques that allow each node to conduct low-complexity computations while asymptotically converging to a global solution~\cite{Nassif2020SPM}, {such as the gossip algorithm~\cite{boyd2006randomized,zhang2020power}, the message passing algorithm~\cite{heusdens2017distributed,heusdens2012distributed}, the primal-dual method of multipliers based on monotone operator theory~\cite{sherson2018derivation,sherson2019distributed} and the distributed adaptive node-specific signal estimation (DANSE) algorithm~\cite{bertrand2010distributed1, bertrand2010distributed2}}.

{Built on these advancements, several distributed algorithms have been proposed for some} front-end speech processing algorithms, such as speech enhancement with distributed Wiener filtering~\cite{bertrand2009robust}, distributed beamforming{\cite{Hu2018PDMM,zhang2019distributed}}, and distributed active noise control~\cite{Dong2020ANC}. {However, few, if any, works consider the problem of distributed speech dereverberation~\cite{pasha2020distributed}.
Therefore, in order to solve the speech dereverberation problem over dispersed sensor networks,} we propose a distributed speech dereverberation based on the WPE method in this work. Specifically, we incorporate DANSE~\cite{bertrand2010distributed1} into the MCLP process{, thereby allocating the computation complexity to each specific node}, and employ the WPE method for parameter determination, {known as a widely-used unsupervised manner in practice, which is different from those supervised methods with training and validation processes~\cite{ml}}.~The proposed framework was validated through both simulated and real distributed scenarios. Experimental results convincingly demonstrate the effectiveness of our proposed approach.

\section{Signal Model and Centralized WPE Algorithm}
Consider an enclosure equipped with a fully interconnected sensor network, consisting of $M$ nodes (microphones). 
\vspace{-5mm}
\subsection{Signal Model}
Let $d(t)$ represent the source speech and $s_i(t)$ represent the reverberant speech received at node $i$, where $t$ denotes the discrete time index. The signal  $s_i(t)$ is related to $d(t)$ through the linear convolution operation with the impulse response $h_i(t)$, which represents the room impulse response (RIR):
 \begin{equation}\label{eq:sigtime}
s_i(t) = h_i(t)\ast d(t)
\end{equation}
where $\ast$ denotes the linear convolution operation.  The signal model in Eq.~\eqref{eq:sigtime} can be approximated in the Short-Time Fourier Transform (STFT) domain:
\begin{equation}\label{eq:sigstft}
\small
{S_i(n,k) = \sum_{\ell=0}^{\tau-1}\underbrace{H_i(\ell,k)D(n-\ell,k)}_{\text{Direct signal}~\&~\text{early reflections}}+\sum_{\ell=\tau}^{L_r-1}\underbrace{H_i(\ell,k)D(n-\ell,k)}_{\text{Late reflections}}}
\end{equation}
where $n$ and $k$ are the time-frame and frequency bin indices respectively. $S_i(n,k)$, $D(n,k)$ and ${H}_{i}(n,k)$ represent the counterparts of $s_i(t)$, $d(t)$ and ${h}_{i}(t)$ in {the} STFT domain. {$L_r$ denotes the length of RIR and $\tau$ denotes the sample index separating the RIR into early and late reverberation reflections~\cite{nakatani2010speech}. }

\vspace{-4mm}
\subsection{Centralized WPE Problem}\label{sec:central}
{As only the late-reverberant components have a detrimental effect on speech intelligibility and quality}, the WPE method~{\cite{nakatani2010speech}} aims to estimate late reflections from delayed observations and subtract them from the reverberant speech signal, resulting in a prediction of the desired speech that includes the direct signal and early reflections. Let ${\mathbf{s}_i(n-{\tau},k)}$ denote a vector of length $L$ on node $i$, representing the STFT domain signal with a delay of {$\tau$} frames
\begin{equation}\label{eq:singleobserve}
{\mathbf{s}_i(n-{\tau},k)} = ({S_i(n-{\tau},k)}, \cdots, {S_i(n-{\tau}-L+1,k)})^{\top}
\end{equation}
{where $(\cdot)^{\top}$ represents vector transpose}. let ${\bw}_i(k)$ denote the prediction weight vector of length $L$.  The process of estimating late reflections and generating the predicted desired speech can be expressed as follows:\begin{equation}\label{eq:wpeproblem}
       \begin{split}
       \hat{D}(n,k) &= S_{\rm ref}(n,k) - \sum_{i=1}^M\ {\bw}^{\rm H}_i(k) \, {{\bs}_i(n-{\tau},k)} \\
                          &= S_{\rm ref}(n,k) - \overline{\bw}^{\rm H}(k) {\overline{\bs}(n-{\tau},k)}
       \end{split}
\end{equation}
where {$(\cdot)^{\rm H}$ represents conjugate transpose and}
\begin{align}
      \overline{\bw}(k) &= \text{col}\{\bw_1(k), \cdots, \bw_M(k)\},\\
      {\overline{\bs}(n-{\tau},k)} &= \text{col}\{{\bs_1(n-{\tau},k)},\cdots, {\bs_M(n-{\tau},k)}\}
\end{align}
are vectors formed by stacking prediction weight vectors and signal vectors of all nodes, respectively, and $S_{\rm ref}(n,k)$ represents the reference {channel randomly chosen from the received signal.}
In order to estimate {the optimal parameters},
a maximum likelihood criterion is employed~{\cite{nakatani2010speech}}, leading to minimize the following cost function:
\begin{equation}\label{eq:wpe}
\mathcal{J}\Big(\overline{\mathbf{w}}(k), \boldsymbol{\sigma}(k)\Big) = \sum_{n=1}^N \frac{|\hat{D}(n,k)|^2}{\sigma(n,k)}+\log\pi\sigma(n,k)
\end{equation}
with ${\boldsymbol{\sigma}}(k) =[\sigma(1,k),\cdots,\sigma(N,k)]^\top$ {representing the Power Spectral Density (PSD) of the desired signal.}
The estimation of $\overline{\mathbf{w}}(k)$ and $\boldsymbol{\sigma}(k)$ can be performed iteratively. Given $\boldsymbol{\sigma}(k)$, $\overline{\mathbf{w}}(k)$ is obtained using the closed-form solution:
\begin{equation}\label{eq:wpew}
\begin{aligned}
\overline{\mathbf{w}}(k) = [Z_{\overline{\mathbf s}}(k)]^{-1}\mathbf{q}_{\overline{\mathbf s}}(k),
\end{aligned}
\end{equation}
where $Z_{\overline{\mathbf s}}(k) = \sum_{n=1}^{N}\frac{{\overline{\mathbf{s}}(n-{\tau},k)}[{\overline{\mathbf{s}}(n-{\tau},k)}]^{\rm H}}{\sigma(n,k)}$ and $\mathbf{q}_{\overline{\mathbf s}}(k) = \sum_{n=1}^{N}\frac{{\overline{\mathbf{s}}(n-{\tau},k)}\hat{D}(n,k)}{\sigma(n,k)}$.


\section{Distributed WPE Based on DANSE}

While the centralized strategy mentioned in Sec.~\ref{sec:central} can effectively seek the optimal filter for speech dereverberation by solving Eq.~\eqref{eq:wpe}, gathering and processing all the observations at each node can be computationally demanding. In this section, we propose an alternative approach, specifically a distributed strategy, where each node has its own independent processing unit. The optimal estimation is achieved through efficient cooperation among the nodes in the network, {i.e., exchanging the compressed data rather than the raw data. With such a strategy, the computation complexity over the whole network can be distributed to each specific node and the communication
bandwidth can be reduced} while maintaining dereverberation performance.

{The DANSE$_1$~\cite{bertrand2010distributed1} algorithm (a specific version of the general DANSE) empowers each node to process locally and ensures the convergence to the global solution. However, it is specifically devised for seeking the solution to the minimum mean-square error problem, while the form of problem  \eqref{eq:wpe}  does not straightforwardly conform with the DANSE$_1$ formulation. Hence, it is necessary to carefully integrate DANSE$_1$ with the vanilla WPE with proper orders of updating the optimization variables.  Recall that the problem~\eqref{eq:wpe} could be} solved iteratively over $\{\overline{\bw}(k)\}_{k=1}^K$ and $\{\boldsymbol{\sigma}(k)\}_{k=1}^K$. When $\{\boldsymbol{\sigma}(k)\}_{k=1}^K$ is fixed, the problem boils down to a (weighted) least-squares problem {with respect to $\{\overline{\bw}(k)\}_{k=1}^K$}, {consistent with the assumption of  the DANSE$_1$ strategy. Moreover, benefiting from the node-specific characteristic of the DANSE$_1$~\cite{bertrand2010distributed1} algorithm, node $i$ could process its raw data and the compressed data from other nodes $\forall j\neq i$ locally.} Therefore, on node $i$, the prediction process now turns to
\begin{equation}\label{eq:dwpeproblem}
\hat{D}_i(n,k)  ~= ~{S_{i}(n,k) \!-\!  \left[\!\begin{array}{c;{2pt/2pt}c}
\bw_{i}^{\rm H}(k)&\tilde{\bw}_i^{\rm H}(k)\\
\end{array}\!\right]
\left[\begin{array}{c}
\mathbf{{s}}_i(n-{\tau},k)\\
\hdashline[2pt/2pt]
\mathbf{{b}}_{i}(n,k)
\end{array}\right]}
\end{equation}
{where $\hat{D}_i(n,k)$ represents the desired signal at node $i$.} In the above expression, {each node $i$ shares a compressed version of its data using a compressor $\bg_{i}(k)$ instead of transmitting the raw data, and $\mathbf{b}_i(n,k)\in\mathbb{C}^{(M-1)\times 1}$} represents the compressed data from other nodes $j \neq i$, such that
\begin{equation}\label{eq:compress}
        \mathbf{b}_i(n,k) = \text{col}\{\, \bg_j^{\rm H}(k)\bs_j(n,k)\,\}_{j\neq i}.
\end{equation}
{$\bw_i(k)\in\mathbb{C}^{L\times 1}$} is the weight vector associated with the local data {$\bs_i(n-\tau,k)\in\mathbb{C}^{L\times 1}$}, and {$\tilde{\bw}_i(k)\in\mathbb{C}^{(M-1)\times 1}$} is the weight vector associated with the received compressed data from nodes $j \neq i$. To ensure the convergence of~{\eqref{eq:dwpeproblem}}, DANSE$_{1}$ {suggests} to set
\begin{equation}\label{eq:com}
        \bg_j(k) = \bw_j(k).
\end{equation}
{In such way, we can see that $\bw_j(k)$ acts as both a compressor and part of the estimator}. Now, the cost function at node $i$ can be expressed by
\begin{equation}\label{eq:newproblem}
\begin{split}
\mathcal{J}_i \Big(\bw(k)_i,\tilde{\bw}_i(k),~&{\boldsymbol{\sigma}_i(k)}\Big)  = \log \pi {\sigma_i(n,k)} \\
 + {\Big|S_{i}(n,k)} -  &{\left[\!\!\begin{array}{c;{2pt/2pt}c}
\bw_i^{\rm H}(k)&\tilde{\bw}_i^{\rm H}(k)\\
\end{array}\!\!\right]
\left[\begin{array}{c}
\mathbf{{s}}_i(n-{\tau},k)\\
\hdashline[2pt/2pt]
\mathbf{{b}}_{i}(n,k)
\end{array}\right]\Big |^2}
\end{split}
\end{equation}
{where each node $i$ uses a local reference $S_{i}(n,k)$ and variance estimation $\sigma_i(n,k)$. }
Rewrite the stacked version of the distributed observations at node $i$ as
\begin{equation}
{\tilde{\textbf{s}}_{i}(n-\tau,k) {=} \left[\begin{array}{c}
\mathbf{{s}}_i(n-\tau,k)\\
\hdashline[2pt/2pt]
\mathbf{{b}}_{i}(n,k)
\end{array}\right].}
\end{equation}
Then, if given $\boldsymbol{\sigma}_i(f)$, a closed form solution of the optimal filter coefficients can be estimated by
\begin{figure*}[!t]
 \centering
      \includegraphics[width=0.75\textwidth]{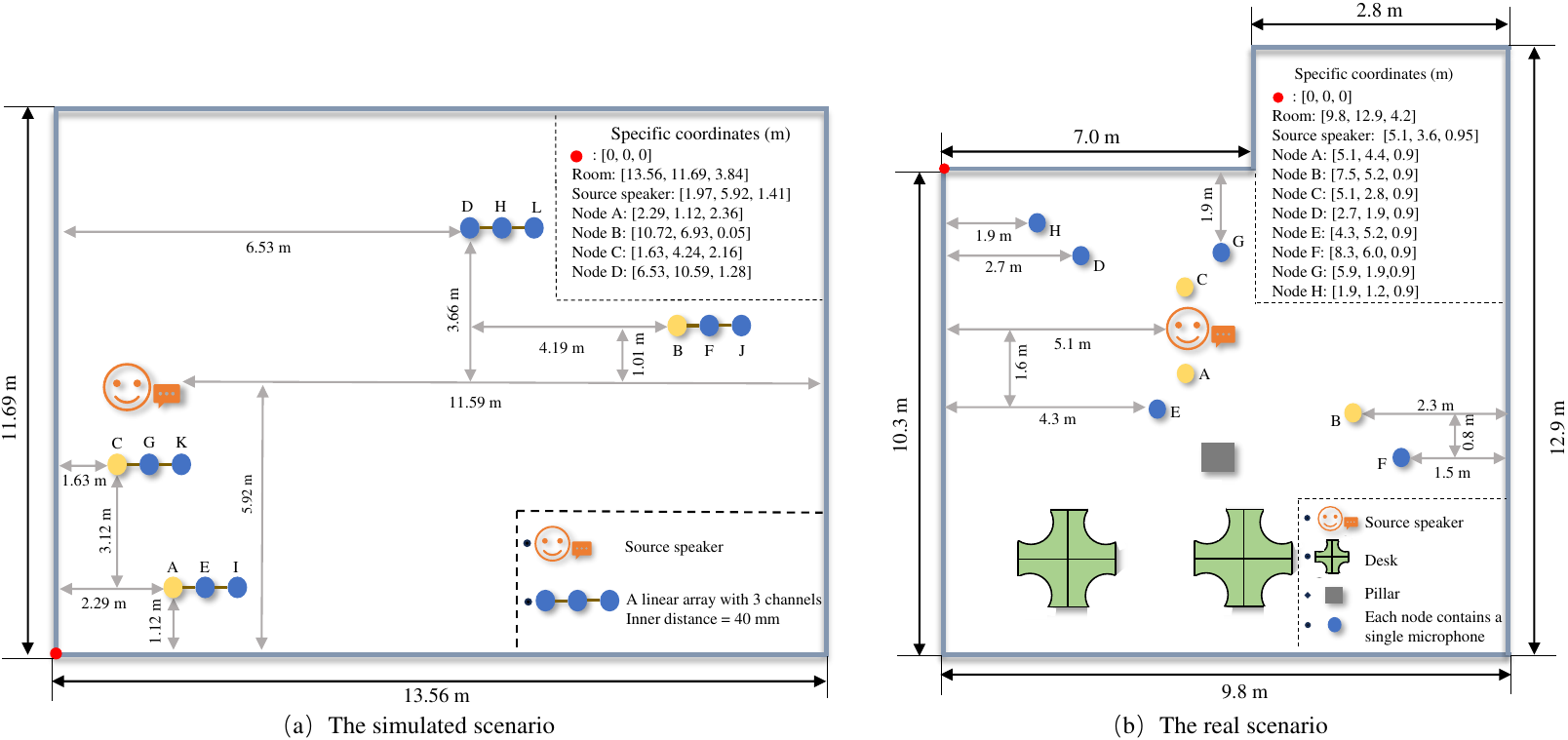}
      \vspace{-0.3cm}
  \caption{The diagram of the experimental scenarios. (a) {The simulated distributed scenario containing 12 nodes. (b) The real distributed scenario containing 8 nodes. The \texttt{Single} method is performed on the node highlighted in yellow.}} 
  \label{fig:room}
  \vspace{-3mm}
\end{figure*}

\begin{equation}\label{eq:c-wpew}
\begin{aligned}
{\left[\begin{array}{c}
\mathbf{{w}}_i(k)\\
\hdashline[2pt/2pt]
\tilde{\bw}_i(k)
\end{array}\right]} = [\mathbf{Z}_{\overline{\mathbf d}_i}(k)]^{-1}\mathbf{q}_{\overline{\mathbf d}_i}(k).
\end{aligned}
\end{equation}
with
\vspace{-0.3cm}
\begin{equation}
\vspace{-0.3cm}
\label{eq:cR}
\mathbf{Z}_{\overline{\mathbf d}_i}(k) = \sum_{n=1}^{N}\frac{{\tilde{\mathbf{s}}_i(n-{\tau},k)}[{\tilde{\mathbf{s}}_i(n-{\tau},k)}]^{\rm H}}{\sigma_i(n,k)}
\end{equation}
and
\vspace{-0.1cm}
\begin{equation}
\label{eq:cq}
\mathbf{q}_{\overline{\mathbf d}_i}(k) = \sum_{n=1}^{N}\frac{{\tilde{\mathbf{s}}_i(n-{\tau},k)}\hat{D}_i(n,k)}{\sigma_i(n,k)}.
\end{equation}
Once $\mathbf{{w}}_i(k)$ and $\tilde{\mathbf{{w}}}_{i}(k)$ are obtained, we can  estimate the PSD of $\hat{D}_i(n,k)$ by:
\begin{equation}\label{eq:c-wpesigmaa}
\sigma_i(n,k) = \text{max} \Big\{ |\hat{D}_i(n,k)|^2, \epsilon \Big\},
\end{equation}
{where $\hat{D}_i(n,k)$ is calculated by \eqref{eq:dwpeproblem}, and $ \epsilon$ is a lower bound that needs to be predefined. }

The overall scheme of the proposed distributed method is {summarized in Algorithm~\ref{alg:dwpe} here, and illustrated in Fig.~S1 in the supplementary material file}.
\begin{algorithm}[t!]
\footnotesize
    \caption{{The proposed framework}}
    \label{alg:dwpe}
    \KwIn {The observed speech signal:~$\mathbf{s}_{i}(n-\tau,k)\in\mathbb{C}^{L\times 1}$.}
    \KwOut { The estimated speech:~$\mathbf{\hat{D}}_i\in \mathbb{C}^{N\times K}$.}
    \textbf{Parameters:}~The time delay: $\tau$; The filter order: $L$; The lower bound:~$\epsilon$; The number of nodes:~$M$; The collaborative decision:~$a$. \\
     \textbf{Initialize:} The filter weights:~${\mathbf{w}}_i(k) = \boldsymbol{0}\in \mathbb{C}^{L\times 1}$ and~$\tilde{\bw}_i(k) = \boldsymbol{0}\in \mathbb{C}^{(M-1)\times 1}$;
    The compressed vector:~$\bg_i(k) = \boldsymbol{0}\in \mathbb{C}^{L\times 1}$; The iteration index:~$\ell = 1$.\\
    \While{ not converged}{
     \For{k = 1 to K}{
     \For{n = 1 to N}{
      Calculate $\sigma_i(n,k)^{(\ell)}$ by \eqref{eq:c-wpesigmaa}\;
    }
      Update $\bw_i(k)^{(\ell)}$ and $\tilde{\bw}_i(k)^{(\ell)}$ by \eqref{eq:c-wpew}\;}
      Update $\mathbf{\hat{D}}^{(\ell)}_i$  by \eqref{eq:dwpeproblem}\;
      \If{$\ell~\%~a = 0$}{
      Update the compressor by \eqref{eq:com}\;
     Compress the local data via $\bg_i(k)^{(\ell)}$\;
     Send the compressed data to other nodes\;}
     $\ell = \ell+1$}
\end{algorithm}

\vspace{-0.4cm}
\section{Experiments}\label{sec:exp}
\vspace{-0.1cm}
\subsection{Experimental Setting}

To evaluate the proposed method, we conducted experiments in both simulated and real environments {with the assumption of a fully-connected network}. In the simulated environment, we constructed an enclosed space with one speaker and four distributed linear microphone arrays~{(each array contains three omnidirectional microphones). Note that we regarded each single microphone as an individual node, thus obtaining a total of 12 nodes}.
{RIRs} were synthesized using the image method~\cite{allen1979image}, and the clean speech (selected from the Wall Street Journal dataset (WSJ0)\cite{garofolo1993csr}) was convolved with these RIRs to generate multi-channel reverberant speech. The reverberation time ({T$_{60}$}) was set to approximately 830 milliseconds (ms). Additionally, we selected a real-world scenario from the Libri-adhoc40 dataset~\cite{guan2021libri}, which was recorded by distributed devices in an office setting {with a pillar and two desks}, with a {T$_{60}$} close to 900 ms. The specific parameters of these scenarios are depicted in {Fig.~\ref{fig:room}}.


We compared the proposed method with two information fusion strategies: a single-channel speech dereverberation method (referred to as \texttt{Single}) and a centralized speech dereverberation method (referred to as \texttt{Centralized}). {The \texttt{Single} method only utilizes the local signal without the information cooperation among different nodes, while} the \texttt{Centralized} method utilizes {the} global information and serves as an upper bound for dereverberation performance. Our proposed method is referred to as \texttt{Distributed}. To quantify the experimental results, we adopt three widely used evaluation measures for speech dereverberation, namely perceptual evaluation of speech quality (PESQ) \cite{torcoli2021objective,rix2001perceptual}, cepstral distance (CD) \cite{kinoshita2016summary} and frequency-weighted segmental signal-to-noise ratio~(F-SNR)~\cite{torcoli2021objective,kinoshita2016summary}. {Here, the \texttt{Signle} method is performed on three selected nodes (highlighted in the yellow circles in Fig. 1).}


\begin{table*}[t!]
\footnotesize
\renewcommand\arraystretch{1.2}
\caption{The results of all comparison methods under both the simulated and real scenarios, {where $T$ denotes the amount of data transmission when calculating the late reverberation for the current frame at each frequency bin.}}\label{table:t1}\vspace{-3mm}
\centering
\setlength{\tabcolsep}{1.4mm}{
\begin{tabular}{c|c|c|c|ccc|ccc|ccc}
\hline
\multicolumn{13}{c}{\cellcolor{gray!30}\textbf{The simulated scenario {(containing 12 nodes in total)}}}\\ \cline{1-13}
\multirow{2}{*}{The number of {nodes}}&\multirow{2}{*}{Method} & \multirow{2}{*}{$L$} &  \multirow{2}{*}{{$T$}} &\multicolumn{3}{c|}{Node A} &\multicolumn{3}{c|}{Node B}&\multicolumn{3}{c}{Node C}\\\cline{5-13}
&&&&PESQ $\uparrow$  &CD $\downarrow$&F-SNR $\uparrow$&PESQ $\uparrow$ &CD$\downarrow$ &F-SNR$\uparrow$&PESQ $\uparrow$ &CD $\downarrow$&F-SNR$\uparrow$\\\cline{1-13}
\textbf{-} &Unprocessed&\textbf{-}&\textbf{-}&1.69&4.81&7.46&1.47&5.27&6.59&2.03&4.05&9.36\\\cline{1-13}
 \textbf{-}&Single&26&\textbf{-}&1.84&4.52&8.09&1.73&4.56&7.72&2.25&3.59&10.38\\\cline{1-13}
6 nodes & Centralized&156&{130}&2.53&3.14&9.72&2.08&3.38&8.71&\textbf{-}&\textbf{-}&\textbf{-}\\\cline{2-13}
(Comprise node A and B)& Distributed&31&{5}&2.27&3.95&8.92&1.83&4.16&8.25&\textbf{-}&\textbf{-}&\textbf{-}\\\cline{1-13}
9 nodes & Centralized &234&{208}&2.52&3.07&9.82&2.13&3.35&8.71&2.60&2.50&12.04\\ \cline{2-13}
(Comprise node A, B and C) & Distributed &34&{8}&2.44&3.68&9.20&1.74&3.99&8.20&2.76&2.72&12.37\\ \cline{1-13}
12 nodes & Centralized&312&{286} &2.53&3.14&9.72&2.08&3.38&8.71&2.58&2.53&11.96\\ \cline{2-13}
 (Comprise node A, B and C)& Distributed &37&{11}&2.56&3.52&9.48&1.89&3.81&8.35&2.88&2.59&12.50\\ \hline
\multicolumn{13}{c}{\textbf{\cellcolor{gray!30}The real scenario {(containing 8 nodes in total)}}}\\ \cline{1-13}
\multirow{2}{*}{The number of {nodes}}&\multirow{2}{*}{Method}&\multirow{2}{*}{$L$} &  \multirow{2}{*}{\cred{$T$}}   &\multicolumn{3}{c|}{Node A} &\multicolumn{3}{c|}{Node B}&\multicolumn{3}{c}{Node C}\\\cline{5-13}
&&&&PESQ $\uparrow$  &CD $\downarrow$&F-SNR$\uparrow$ &PESQ $\uparrow$ &CD$\downarrow$ &F-SNR$\uparrow$&PESQ $\uparrow$ &CD $\downarrow$&F-SNR$\uparrow$\\\cline{1-13}
\textbf{-} &Unprocessed&\textbf{-}&\textbf{-}&2.56&4.14&8.93&1.58&5.59&4.34&1.95&5.86&5.03\\\cline{1-13}
\textbf{-} &Single&40&\textbf{-}&2.90&3.47&10.25&1.64&5.15&4.95&2.09&5.33&5.72\\\cline{1-13}
4 nodes  & Centralized&160&{120}&3.17&3.34&11.46&2.28&4.65&6.82&\textbf{-}&\textbf{-}&\textbf{-}\\\cline{2-13}
(Comprise node A and B)& Distributed&43&{3}&2.94&3.42&10.48&1.72&5.09&5.05&\textbf{-}&\textbf{-}&\textbf{-}\\\cline{1-13}
6 nodes  & Centralized& 240&{200}&3.24&3.36&11.43&2.37&4.81&6.94&2.67&4.79&7.10\\ \cline{2-13}
(Comprise node A, B and C) & Distributed &45&{5}&2.97&3.38&10.66&1.75&5.05&5.13&2.20&5.22&5.96\\ \cline{1-13}
8 nodes  & Centralized &320&{280}&3.25&3.40&11.50&2.39&4.83&7.06&2.68&4.89&7.05\\ \cline{2-13}
(Comprise node A, B and C) & Distributed& 47&{7}&3.01&3.36&10.81&1.81&5.03&5.20&2.23&5.17&6.09\\
\hline
\end{tabular}}
\vspace{-5mm}
\end{table*}

\vspace{-4.4mm}

\subsection{Implementation Details}
All comparison methods were implemented in the STFT domain with a Hann window, {with the frame length of 32 ms and 75\% overlapping.} The filter order $L$ was set to 26 for the simulated scenario and 40 for the real scenario. We conducted experiments on ${\tau}$ ranging from 2 to 6 and determined the optimal values of 4 for the simulated scenario and 5 for the real scenario. In the proposed method, {we set $a=2$ to reduce the possibility of receiving error data from other nodes. Additionally, we used the generalized cross-correlation with phase transform function to synchronize all the signals.}
\vspace{-4mm}
\subsection{Result Discussion}
\vspace{-2mm}
\begin{table}[t!]
\footnotesize
\renewcommand\arraystretch{1.2}
\caption{ {The values of $\beta_{(\cdot)}$
with different number of nodes.}}\label{table:t2}\vspace{-4mm}\vskip 5pt
\centering
\setlength{\tabcolsep}{1.5mm}{
\begin{tabular}{c|c|c|c|c|c|c}
\toprule
\multirow{2}{*}{{The number of nodes}}& \multicolumn{3}{c|}{The simulated scenario}& \multicolumn{3}{c}{The real scenario}\\ \cline{2-7}
& 6& 9& 12& 4& 6& 8\\ \cline{1-7}
$\beta_{(\times)}~\text{in}~{\eqref{eq:cR}}~\&~\eqref{eq:cq}$  & 0.042& 0.022&0.015& 0.076& 0.037&0.023\\ \cline{1-7}
$\beta_{(\div)}~\text{in}~{\eqref{eq:cR}}~\&~\eqref{eq:cq}$  & 0.205& 0.150&0.122& 0.275& 0.192&0.150\\ \cline{1-7}
$\beta_{(/)}~\text{in}~{\eqref{eq:c-wpew}}$   & 0.009& 0.003&0.002& 0.021& 0.007&0.003\\
\bottomrule
\end{tabular}}\vspace{-1mm}
\end{table}
\begin{figure}[!t]
 \centering
      \includegraphics[width=0.4\textwidth]{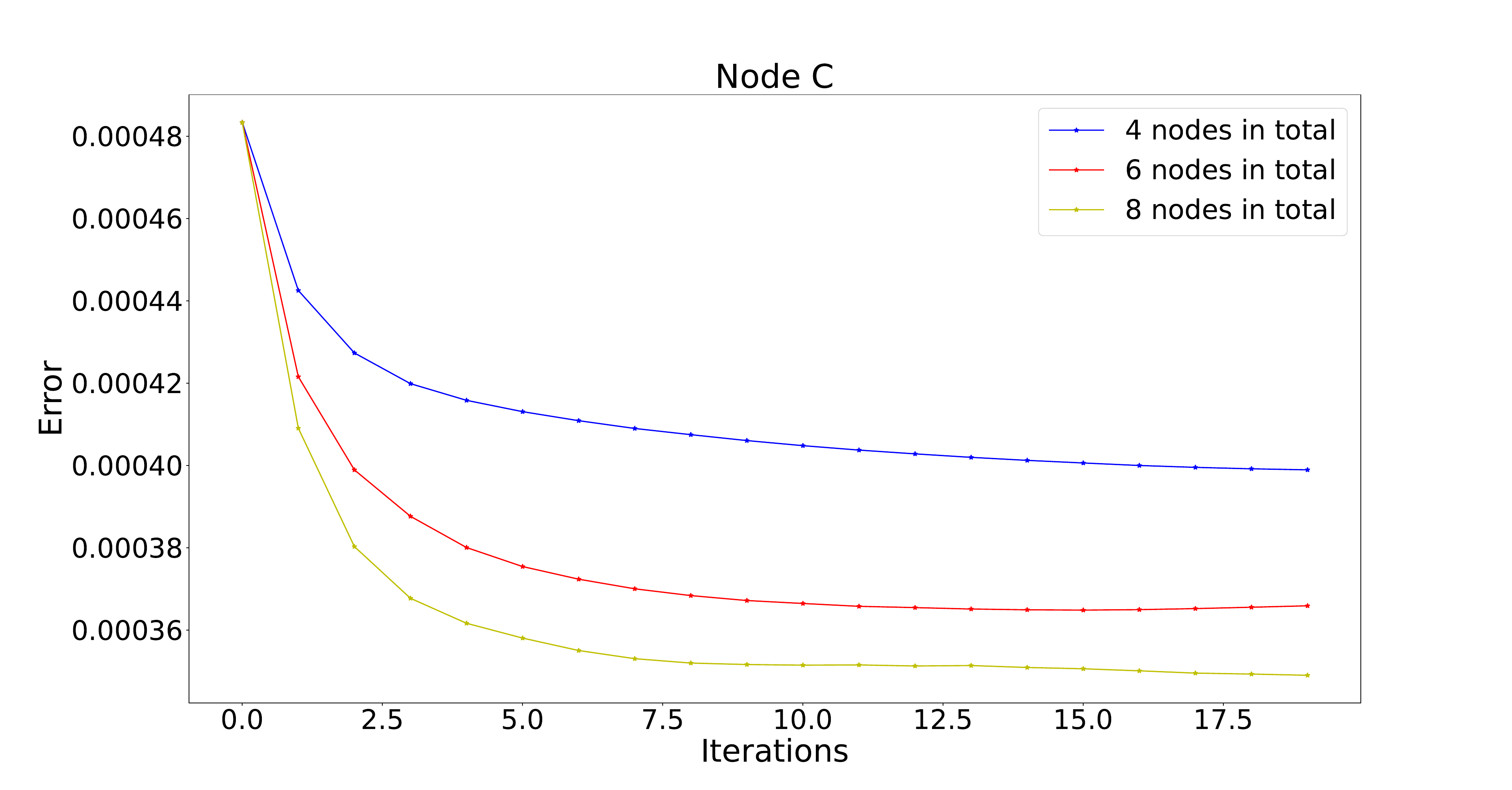}
  \caption{{The error convergence curves of the proposed method in the real scenario, where the vertical axis is defined in the supplemental material.}}
  \label{fig:conv}\vspace{-7mm}
\end{figure}
Table~\ref{table:t1} presents the results of three selected nodes (Node A, B, and C) in both simulated and real scenarios. It can be observed that the \texttt{Distributed} method consistently outperforms the \texttt{Single} method in all experiments, as expected. For example, in the simulated scenario with 6 nodes fusion, the \texttt{Distributed} method achieves improvements of 0.43 (PESQ), 0.57 (CD), and 0.83 (F-SNR) compared to the \texttt{Single} method, while only slightly increasing the filter order by 5. Furthermore, we made an interesting observation that as the number of nodes increases, the performance of the \texttt{Centralized} method degrades, while the \texttt{Distributed} method still maintains its advantages. In several conditions under the simulated scenario, the \texttt{Distributed} method even outperforms the \texttt{Centralized} method in terms of PESQ and F-SNR. This can be attributed to the fact that a larger matrix of observations in the \texttt{Centralized} method is more prone to errors in \eqref{eq:wpew}, whereas the proposed \texttt{Distributed} method compensates for this issue through {the} compressed information transmission.

{To demonstrate the bandwidth reduction of the \texttt{Distributed} method, we calculated the amount of data transmission~(reported in Table.~\ref{table:t1}) and concluded that the \texttt{Distributed} method could reduce the amount of data transmission by 96.15~\% and 97.5~\% in the simulated and real scenarios, respectively.} To demonstrate the computational efficiency of the \texttt{Distributed} method, we compared the computational complexity\footnote{{The calculation details of the computational complexity and the definition of the reduction factors ($\beta_{(\cdot)}$) could be found in Table S1 in the supplemental material file.}} of it with that of the \texttt{Centralized} method, {reported in Table~\ref{table:t2}}.
We can observe that $\beta_{(\cdot)}$ is significantly smaller than 1 and decrease with the number of nodes, {indicating the computational efficiency of} the \texttt{Distributed} method. {Additionally, we present the error evolution in Fig.~\ref{fig:conv} and Fig.~S2 in the supplemental file, empirically showing the convergence of the method in terms of different number of nodes. }
\vspace{-3mm}
\section{Conclusions}\label{sec:conclu}
\vspace{-1mm}
In this paper, we proposed a method for {the} distributed speech dereverberation algorithm to adapt {to} the scenario with dispersed nodes of limited computational capacity. Our approach leveraged the DANSE strategy to distribute the computation of the WPE method. Each node in the distributed network performed parallel processing and cooperated with other nodes to achieve speech dereverberation. Future work can explore the extension of the proposed method to handle more general cases where each node consists of an array of microphones. Additionally, investigating network clustering and node selection techniques would be beneficial for efficient processing in large-scale distributed networks.

\bibliographystyle{IEEEbib}
\bibliography{refs}

\end{document}